\documentclass[submission,copyright,creativecommons]{eptcs}
\providecommand{\event}{QAPL 2019} % Name of the event you are submitting to

\usepackage{underscore}           % Only needed if you use pdflatex.
\usepackage{tikz}
\usepackage{stmaryrd}
\usepackage{amsmath}
\usepackage{amssymb}
\usepackage{amsfonts}
\usepackage{amsthm}
\usepackage{thmtools}

\newcommand{\dist}{\mathcal{D}}
\newcommand{\subdist}{\dist_{\leq}}
\newcommand{\Exp}[1]{\mathit{Exp}\!\left[#1\right]}
\newcommand{\wrt}[1]{\mathop{\mathrm{d}#1}}

\newcommand{\paths}{\Pi}

\newcommand{\cylinder}{\mathfrak{C}}
\newcommand{\prob}{\mathbb{P}}
\newcommand{\ft}{\preceq}
\newcommand{\simul}{\precsim}
\newcommand{\eqft}{\equiv}

\newcommand{\comp}[1]{\mathbin{\|_{#1}}}
\newcommand{\mon}[2]{\mathbin{\lessapprox^{#1}_{#2}}}
\newcommand{\smon}[2]{\mathbin{\leqq^{#1}_{#2}}}

\declaretheorem[numberwithin=section,style=plain]{theorem}
\declaretheorem[numberlike=theorem,style=definition,qed=$\blacktriangle$]{definition} 
\declaretheorem[numberlike=theorem,style=definition,qed=$\blacklozenge$]{example}

\declaretheorem[numberlike=theorem,style=plain]{lemma}
\declaretheorem[numberlike=theorem,style=plain]{proposition}
\declaretheorem[numberlike=theorem,style=plain]{corollary}

\usetikzlibrary{arrows,automata,positioning,calc}
\tikzset{
    every node/.style={
        font={\fontsize{8pt}{12}\selectfont}
    }
}

\title{A Faster-Than Relation for Semi-Markov Decision Processes}
\author{Mathias Ruggaard Pedersen \quad Giorgio Bacci \quad Kim Guldstrand Larsen
\institute{Department of Computer Science\\
Aalborg University\\
Denmark}
\email{\{mrp,grbacci,kgl\}@cs.aau.dk}
}
\def\titlerunning{A Faster-Than Relation for SMDPs}
\def\authorrunning{M.R. Pedersen, G. Bacci, \& K.G. Larsen}
\begin{document}
\maketitle

%% Abstract
\begin{abstract}
  When modeling concurrent or cyber-physical systems,
  non-functional requirements such as time
  are important to consider. In order to improve the timing aspects of a model,
  it is necessary to have some notion of what it means
  for a process to be faster than another,
  which can guide the stepwise refinement of the model.
  To this end we study a \emph{faster-than relation} for
  semi-Markov decision processes
  and compare it to standard notions for relating systems.
  We consider the compositional aspects of this relation, 
  and show that the faster-than relation is not a precongruence
  with respect to parallel composition, hence 
  giving rise to so-called parallel timing anomalies.
  We take the first steps toward understanding this problem
  by identifying decidable conditions sufficient to avoid parallel timing anomalies 
  in the absence of non-determinism.
\end{abstract}

%% Introduction
\section{Introduction}
Timing aspects are important when considering real-time or cyber-physical systems.
For example, they are of interest in real-time embedded systems when one wants to verify the worst-case execution time for guaranteeing minimal system performance
or in safety-critical systems when one needs to ensure that unavoidable rigid deadlines will always be met~\cite{lee2008}. 

Semi-Markov decision processes are continuous-time Markov decision processes where the residence-time on states is governed by generic distributions on the positive real line.
These systems have been extensively used to model real-time cyber-physical systems~\cite{SSR15,TS16}. 

For reasoning about timing aspects it is important to understand what it formally means 
for a real-time or cyber-physical system to operate faster than another. 
To this end we consider a notion of \emph{faster-than relation} for semi-Markov decision processes.
The definition of faster-than relation we propose in this paper is a reactive version of an 
analogous notion of faster-than relation previously introduced in~\cite{PFBLM18} for the case 
of generative systems.
According to this relation, a semi-Markov decision process is faster than another one when
it reacts to any sequence of inputs with equal or higher probability than the slower process, within the same time bound.

Often, complex cyber-physical systems are organised as concurrent systems 
of multiple components running in parallel and interacting with each other.
Such systems are better analysed \emph{compositionally}, that is, by breaking them into smaller 
components that are more easily examined~\cite{CLM89}.
However, it is not always the case that an analysis on the components
carries over to the full composite system. A well known example of this, occurring in
real-time systems such as scheduling for processors~\cite{cassez2012,lundqvist1999}, are 
\emph{timing anomalies}, that is, when locally faster behaviour leads to a globally slower behaviour~\cite{kirner2009}.

\begin{figure}
  \center
  \begin{tikzpicture}
    % Context
    \draw (0,0) -- (3,0) -- (3,3) -- (0,3) -- (0,1.8);
    \draw (0,1.2) -- (0,0);
    \draw (0,1.8) .. controls  (0.7,2.0) and (0.7,1.0) .. (0,1.2);
    \node[align=center] at (1.5,1.5) {Context \\ $W$};
    
    % Component 1
    \draw (0,0.6) -- (-1.6,0.6) -- (-1.6,2.4) -- (0,2.4);
    \node[align=center] at (-0.8,1.6) {Component \\ $V$ \\ slow};
    
    % Component 2
    \draw (-5,0.6) -- (-6.6,0.6) -- (-6.6,2.4) -- (-5,2.4);
    \draw (-5,2.4) -- (-5,1.8);
    \draw (-5,1.2) -- (-5,0.6);
    \draw (-5,1.8) .. controls (-4.3,2.0) and (-4.3,1.0) .. (-5,1.2);
    \node[align=center] at (-5.8,1.6) {Component \\ $U$ \\ fast};
    
  \end{tikzpicture}
  \caption{The context $W$ operates in parallel with the component $V$.
           If the component $U$ is faster than $V$, then if we replace $V$ with $U$,
           we would expect the overall behaviour to also be faster.}
  \label{fig:ft}
\end{figure}
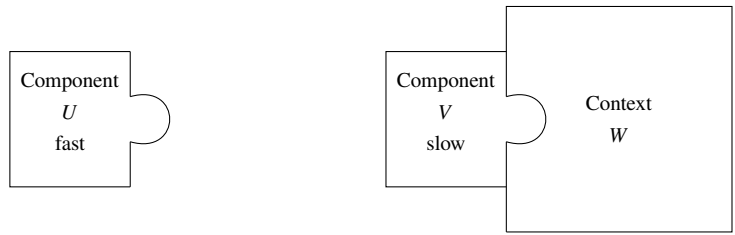

In this paper we study the compositional aspects of the faster-than relation for semi-Markov
decision processes. We are interested in the situation depicted in Figure~\ref{fig:ft}
where we have a composite system consisting of a context $W$ and a component $V$,
and we want to understand what happens when we replace $V$ with another component $U$ that
is faster than $V$. We consider common notions of parallel composition, and show that 
timing anomalies can occur using our faster-than relation, even in the absence of
non-determinism. This shows that
timing anomalies are not caused by non-determinism, but arise from 
the linear timing behaviour of processes.

As a first step toward recovering the compositional reasoning for the faster-than relation,
we identify a condition, called \emph{monotonicity}, to be checked on the components taking
part to the parallel composition which guarantees that timing anomalies cannot occur. 
Our condition is parametric in the parallel composition operator and applies to 
all the notions of parallel composition considered in this paper. 

Presently, we do not know whether the monotonicity condition is decidable or not, however we introduce 
another condition, called \emph{strong monotonicity}, still parametric on the parallel composition operator,
that we show to be decidable. This decidability result is remarkable for a series of reasons.
In~\cite{PFBLM18}, the faster-than relation over generic semi-Markov processes was proven to be 
undecidable and \emph{Positivity-hard} if restricted to processes without 
non-determinism. The Positivity-hardness relates the faster-than problem to the Skolem problem, an important problem in number theory, whose decidability status has been an open for at least 
80 years~\cite{OW14,AAOW15}.
In the light of these results, it is not surprising that strong monotonicity condition only applies 
to processes with no non-determinism.

\textbf{Related Work.}
The notion of a faster-than relation has been studied in many different contexts throughout the literature.
The work most closely related to ours is that of Pedersen et al. \cite{PFBLM18},
which considers a generative version of the faster-than relation,
whereas we study the reactive version.
The focus of \cite{PFBLM18} is on decidability issues,
and the faster-than relation is proved undecidable,
and the proof can easily be extended to also show that the reactive version is undecidable.
However, positive results are also given
in the form of an approximation algorithm,
and a decidability result for unambiguous processes.
Baier et al. \cite{BKHW05} define, among other relations,
a simulation relation for continuous-time Markov chains
which can be interpreted as a faster-than relation,
and study its logical characterisation.
However, none of these works consider compositional aspects.

For process algebras, discrete-time faster-than relations have been defined for variations of Milner's CCS,
and shown to be precongruences with respect to parallel composition \cite{corradini1995,LV01,MT91,satoh1994}.
L{\"u}ttgen and Vogler \cite{luttgen2006} attempt to unify some of these process algebraic approaches
and also consider the issue of parallel timing anomalies.
For Petri nets, Vogler \cite{vogler1995a,vogler1995b} considers a testing preorder as a faster-than relation
and shows that this is a precongruence with respect to parallel composition.
Geilen et al. \cite{geilen2011} introduces a refinement principle for timed actor interfaces
under the slogan ``the earlier, the better'', which can also be seen as an example of a faster-than relation.

Work on timing anomalies date back to at least 1969 \cite{graham1969},
but the most influential paper in the area is probably that of Lundqvist and Stenstr\"{o}m \cite{lundqvist1999},
which shows that timing anomalies can occur in dynamically scheduled processors.
More recent work has focused on compositional aspects \cite{kirner2009}
and defining timing anomalies formally, using transition systems as the formalism \cite{cassez2012,reineke2006}.

%% Preliminaries
\tikzset{->, >=latex}

\section{Notation and Preliminaries}
In this section we fix some notation and recall concepts
that are used throughout the rest of the paper.
Let $\mathbb{N}$ denote the natural numbers and
$\mathbb{R}_{\geq 0}$ denote the non-negative real numbers,
which we equip with the standard Borel $\sigma$-algebra $\mathbb{B}$.
For any set $X$, $\dist(X)$ denotes the set of probability measures on $X$,
and $\subdist(X)$ the set of subprobability measures on $X$.
For an element $x \in X$,
we will use $\delta_x$ to denote the Dirac measure at $x$
defined as $\delta_x(E) = 1$ if $x \in E$ and $\delta_x(E) = 0$ otherwise.
We fix a non-empty, countable set $L$ of \emph{labels} or \emph{actions}
and equip them with the discrete $\sigma$-algebra $\Sigma_L$.

For a probability measure $\mu \in \dist(\mathbb{R}_{\geq 0})$,
we denote by $F_\mu$ its \emph{cumulative distribution function (CDF)} defined as $F_{\mu}(t) = \mu([0,t])$, for all $t \in \mathbb{R}_{\geq 0}$.
We will denote by $\Exp{\theta}$ the CDF of an exponential distribution with rate $\theta > 0$.
The \emph{convolution} of two probability measures $\mu, \nu \in \dist(\mathbb{R}_{\geq 0})$, 
denoted by $\mu * \nu$, is the probability measure on $\mathbb{R}_{\geq 0}$ given by $(\mu * \nu)(B) = \int_{-\infty}^{\infty} \nu(B - x) \;\mu(\wrt{x})$,
for all $B \in \mathbb{B}$ \cite{billingsley1995}.
Convolution is associative, i.e., $\mu * (\nu * \eta) = (\mu * \nu) * \eta$, and commutative, i.e., $\mu * \nu = \nu * \mu$. 

%% Semi-Markov decision processes
\section{Semi-Markov Decision Processes}
In this section we recall the definition of semi-Markov decision processes.
\begin{definition}
  A \emph{semi-Markov decision process (SMDP)} is a tuple
  $M = (S, s_0, \tau,\rho)$ where
  \begin{itemize}
    \item $S$ is a non-empty, countable set of \emph{states},
    \item $s_0 \in S$ is a distinguished \emph{start state},
    \item $\tau : S \times L \rightarrow \subdist(S)$ is a \emph{transition probability function}, and
    \item $\rho : S \rightarrow \dist(\mathbb{R}_{\geq 0})$ is a \emph{residence-time probability function}. \qedhere
  \end{itemize}
\end{definition}

The operational behaviour of an SMDP $M = (S,s_0,\tau,\rho)$ is as follows.
Starting from $s_0$, the process reacts to an external input $a \in L$ provided by the environment 
by changing its state to $s' \in S$ within time $t \in \mathbb{R}_{\geq 0}$ with probability
$\tau(s_0,a)(s') \cdot \rho(s_0)([0, t])$.
The process then continues with the new state $s'$ in place of $s_0$.

Notice that Markov decision processes are a special case of SMDPs
where for all $s \in S$, $\rho(s) = \delta_0$ (i.e. transitions happen instantaneously), and that
continuous-time Markov decision processes are also a special case of SMDPs
where, for all states $s \in S$, $F_{\rho(s)} = \Exp{\theta_s}$ for some rate $\theta_s \in \mathbb{R}_{\geq 0}$.

The executions of an SMDP $M = (S,s_0,\tau,\rho)$ are infinite timed transition sequences of the form
\[\pi = (a_1,t_1,s_1)(a_2,t_2,s_2)\dots \in (L \times \mathbb{R}_{\geq 0} \times S)^\omega,\]
representing the fact that $M$ waited $t_i$ time units after the action $a_i$ was input and then moved to the state $s_i$.
We will refer to executions of an SMDP as \emph{paths}.
We denote by $(\paths(M), \Sigma)$ the \emph{measurable space of paths}, where 
$\Sigma$ is the smallest $\sigma$-algebra generated by the cylinders of the form
$L_1 \times R_1 \times S_0, \dots, L_n \times R_n \times S_n$
for $L_i \in 2^L$, $R_i \in \mathbb{B}$, and $S_i \in 2^S$.
Given a path $\pi$ and $i \in \mathbb{N}$, we let
$\pi |_i = (a_1, t_1, s_1) \dots (a_i, t_i, s_i)$
be the prefix up to $i$ of $\pi$.
We let $\paths(M)$ denote the set of all timed action paths in $M$,
and denote by
\[\paths_n(M) = \{ \pi |_i \mid \pi \in \paths(M)\}\]
the set of all prefixes of length $n$.

In this paper we assume that external choices are resolved by means of memoryless stochastic schedulers, 
however all the results we present still hold for schedulers that can remember actions and states.

\begin{definition}
  Given an SMDP $M = (S, s_0, \tau, \rho)$, a \emph{scheduler} for $M$ is a function $\sigma : S \rightarrow \dist(L)$ that assigns to each state a probability distribution over action labels.
\end{definition}

We will use the notation $\tau^\sigma(s,a)(s')$
as shorthand for $\tau(s,a)(s') \cdot \sigma(s)(a)$ to denote the probability of moving from state $s$ to $s'$
under the stochastic choice of $a$ given by $\sigma$.

Next we recall the standard construction of the measurable space of paths. 
A \emph{cylinder set} of rank $n \geq 1$ is the set of all paths whose $n$-th prefix is contained in a common subset $E \subseteq \paths_n(M)$,
and is given by
\[\cylinder(E) = \{ \pi \in \paths(M) \mid \pi |_n \in E \}.\]
It will be convenient to denote \emph{rectangular cylinders} 
of the form
\[\cylinder(L_1 \times R_1 \times S_1 \times \dots \times L_n \times R_n \times S_n),\]
for $L_i \subseteq L$, $R_i \subseteq \mathbb R_{\geq 0}$, and $S_i \subseteq S$ as
\[\cylinder(L_1 \dots L_n, R_1 \dots R_n, S_1 \dots S_n).\]
We denote by $(\paths(M), \Sigma)$ the \emph{measurable space of paths},
where $\Sigma$ is the smallest $\sigma$-algebra generated by the measurable cylinders of the form
\[\cylinder(L_1 \dots L_n, R_1 \dots R_n, S_1 \dots S_n)\]
for $L_i \in 2^L$, $R_i \in \mathbb{B}$, $S_i \in 2^S$.
Given an SMDP $M$ and a scheduler $\sigma$ for it,
we define inductively the usual probability on cylinders as follows.

\begin{definition}\label{def:prob}
  Let $M = (S,s_0,\tau,\rho)$ be an SMDP. Given a scheduler $\sigma$ for M and a state $s \in S$,
  $\prob^\sigma_M(s)$ is defined as the unique (sub)probability measure\footnote{Existence and uniqueness is guaranteed by the Hahn-Kolmogorov theorem \cite{tao2013}.} on
  $(\paths(M), \Sigma)$ such that for all $L_i \in 2^L$, $R_i \in \mathbb{B}$, and $S_i \in 2^S$ 
  with $1 \leq i \leq n$, we have
  \[\prob_M^\sigma(s)(\cylinder(L_1, R_1, S_1)) = \rho(s)(R_1) \cdot \sum_{a \in L_1} \sum_{s' \in S_1} \tau^\sigma(s,a)(s')\]
  and
  \begin{align*}
    &\phantom{{}={}}\prob_M^{\sigma}(s)(\cylinder(L_1 \dots L_n, R_1 \dots R_n, S_1 \dots S_n)) \\
    &= \rho(s)(R_1) \cdot \sum_{a \in L_1} \sum_{s' \in S_1} \tau^\sigma(s,a)(s') \cdot \prob_M^\sigma(s')(\cylinder(L_2 \dots L_n, R_2 \dots R_n, S_2 \dots S_n)). \qedhere
  \end{align*}
\end{definition}

For example, $\prob_M^\sigma(s)(\{a\} \{b\}, [0,x] [0,y], \{s'\} \{s''\})$
is the probability of doing an $a$-transition from $s$ to $s'$ within $x$ time units
and then doing a $b$-transition from $s'$ to $s''$ within $y$ time units.

%% Faster-than relation
\section{A Faster-Than Relation}
Intuitively, a process $U$ is faster than $V$ if it is able to execute any sequence of actions $a_1, \dots, a_n$ in less time than $V$.
Since our systems are probabilistic, we need to consider the probability of their execution within any time bound.

\begin{figure}
  \centering
  \hfill
  \begin{tikzpicture}
    % Nodes
    \node[state, circle split] (0) {$\mu$ \nodepart{lower} $u_0$};
    \node[state, circle split] (1) [right = 1.5cm of 0]{$\nu$ \nodepart{lower} $u_1$};
    \node[state, circle split] (8) [right = 1.5cm of 1]{$\eta$ \nodepart{lower} $u_2$};
    
    \node[xshift=-0.5cm] at (0.west) {\large{$U$}};
    
    % Edges
    \path[thick] (0) edge [above] node {$(1,a)$} (1);
    \path[thick] (1) edge [above] node {$(1,a)$} (8);
    \path[thick, loop right] (8) edge [right] node {$(1,a)$} (8);
    
    % Nodes
    \node[state, circle split] (2) [below = 0.4cm of 0] {$\nu$ \nodepart{lower} $v_0$};
    \node[state, circle split] (3) [right = 1.5cm of 2] {$\mu$ \nodepart{lower} $v_1$};
    \node[state, circle split] (9) [right = 1.5cm of 3] {$\eta$ \nodepart{lower} $v_2$};
    
    \node[xshift=-0.5cm] at (2.west) {\large{$V$}};
    
    % Edges
    \path[thick] (2) edge [above] node {$(1,a)$} (3);
    \path[thick] (3) edge [above] node {$(1,a)$} (9);
    \path[thick, loop right] (9) edge [right] node {$(1,a)$} (9);
  \end{tikzpicture}
  \hfill \
  \caption{If $F_{\mu}(t) \geq F_{\nu}(t)$ for all $t$, then $U$ is faster than $V$ in the first states, and after that their probabilities are the same,
           so $U$ is faster than $V$.}
  \label{fig:faster-than-c}
\end{figure}
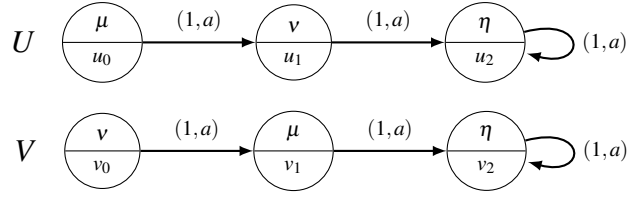

Consider the two simple SMDPs $U$ and $V$ in Figure~\ref{fig:faster-than-c} 
with just a single transition label and initial states $u_0$ and $v_0$, respectively.
Here $\mu,\nu,\eta$ are arbitrary probability measures on $\mathbb{R}_{\geq 0}$, representing
the residence-time distributions at each state.
An arrow with label $(p,a)$ means that when $a$ is chosen as the action,
then the SMDP takes the transition given by the arrow with probability $p$.
The only finite sequences of actions that can be executed in these SMDPs are of the form $a^n$ for $n > 0$.

For $U$ to be faster than $V$, it should be the case that
for any time $t$ and any 
scheduler $\sigma$ for $V$, we must be able to find a scheduler $\sigma'$ for $U$
which allows $U$ to execute
any sequence of actions within time $t$ with higher or equal probability than $V$.
Hence, the type of events on which we want to focus are the following.

\begin{definition}\label{def:cylinder}
  For any finite sequence of actions $a_1, \dots, a_n$,
  and $t \in \mathbb{R}_{\geq 0}$, we say that
  \[\cylinder(a_1 \dots a_n, t) = \{ (b_1,t_1,s_1)(b_2,t_2,s_2)\dots \in \Pi(M) \mid \forall 1 \leq i \leq n,\; b_i = a_i \text{ and } \textstyle\sum_{j = 1}^n t_i \leq t \}\]
  is a \emph{time-bounded cylinder}.
  The \emph{length} of a time-bounded cylinder is the length
  of the sequence of actions in the cylinder.
\end{definition}

Note that $\cylinder(a_1 \dots a_n, t)$ is measurable in $(\paths(M), \Sigma)$,
since
\[f : \paths_n(M) \rightarrow L^n \times \mathbb{R}_{\geq 0}^n \times S^n\]
given by
\[f((o_1,t_1,s_1), \dots, (o_n,t_n,s_n)) = ((o_1, \dots, o_n), (t_1, \dots, t_n), (s_1, \dots, s_n))\]
and
\[\texttt{res}_n : \paths(M) \rightarrow \paths_n(M)\]
given by
\[\texttt{res}_n(\pi) = \pi|_n\]
are both measurable,
and hence
\[(f \circ \texttt{res}_n)^{-1}(S^n \times \{(a_1, \dots, a_n)\} \times B^n_t) = \cylinder(a_1 \dots a_n, t)\]
is measurable,
where $B^n_t = \{(r_1, \dots, r_n) \in \mathbb{R}^n_{\geq 0} \mid \sum_{i = 1}^n r_i \leq t\}$.

\begin{example}
  The time-bounded cylinder $\cylinder(aa,2)$ denotes the set of all paths
  where the first two output labels are both $a$'s,
  and the first two steps of the path are completed within $2$ time units.
\end{example}

For the rest of the paper, we fix three SMDPs
$M = (S, s_0, \tau, \rho)$, $U = (S_U, u_0, \tau_U, \rho_U)$, and $V = (S_V, v_0, \tau_V, \rho_V)$.
Now we are ready to define what it means for an SMDP to be ``faster than'' another one.

\begin{definition}[Faster-than]
  We say that $U$ is \emph{faster than} $V$, written $U \ft V$, if
  for all schedulers $\sigma$ for $V$, time bounds $t$, and sequences of actions $a_1 \dots a_n$,
  there exists a scheduler $\sigma'$ for $U$ and time bound $t' \leq t$, such that 
  $\prob^{\sigma'}(u_0)(\cylinder(a_1 \dots a_n, t')) \geq \prob^{\sigma}(v_0)(\cylinder(a_1 \dots a_n, t))$.
\end{definition}

Clearly, the faster-than relation $\ft$ is a preorder.
The following proposition gives a characterisation of
the faster-than relation that is often easier to work with.

\begin{proposition}\label{prop:alternative}
  $U \ft V$ if and only if for all schedulers $\sigma$ for $V$
  there exists a scheduler $\sigma'$ for $U$ such that
  $\prob^{\sigma'}(u_0)(C) \geq \prob^{\sigma}(v_0)(C)$,
  for all time-bounded cylinders $C$.
\end{proposition}
%\begin{proof}%[Proof of Proposition \ref{prop:alternative}]
%  Clearly, if for all schedulers $\sigma$ for $V$ there exists a scheduler $\sigma'$ for $U$ such that
%  $\prob^{\sigma}(u_0)(C) \geq \prob^{\sigma'}(v_0)(C)$
%  for all time-bounded cylinders $C$, then $U \ft V$ by taking $C' = C$.
%  If $U \ft V$, then consider an arbitrary scheduler $\sigma$,
%  and time-bounded cylinder $C = \cylinder(a_1 \dots a_n, t)$.
%  There exists a scheduler $\sigma'$ and
%  $t' \in \mathbb{R}_{\geq 0}$ such that $t \geq t'$ and
%  \[\prob^{\sigma'}(u_0)(a_1 \dots a_n, t') \geq \prob^{\sigma}(v_0)(a_1 \dots a_n, t).\]
%  By monotonicity, $t \geq t'$ implies that
%  \[\prob^{\sigma'}(u_0)(a_1 \dots a_n, t) \geq \prob^{\sigma'}(u_0)(a_1 \dots a_n, t'),\]
%  and hence $\prob^{\sigma'}(u_0)(C) \geq \prob^{\sigma}(v_0)(C)$.
%\end{proof}

Before showing an example of an SMDP being faster than another one, we provide an analytic solution
for computing the probability over time-bounded cylinders in terms of convolutions of the residence time distributions.

\begin{proposition}\label{prop:hyp}
  For any scheduler $\sigma$ for $M$, and $s \in S$ we have
  \begin{align*}
    \prob^\sigma(s)(\cylinder(a_1 \dots a_n, t)) 
    =\sum_{s_1, \dots, s_n \in S} \tau^\sigma(s,a_1)(s_1) \cdots \tau^\sigma(s_{n-1},a_n)(s_n) \cdot (\rho(s) * \rho(s_1) * \dots * \rho(s_{n-1}))([0,t]) .
  \end{align*}
\end{proposition}
%\begin{proof}%[Proof of Proposition \ref{prop:hyp}]
%  By Corollary \ref{cor:prodmeasure}, we know that
%  \begin{align*}
%    &\phantom{{}={}} \prob^\sigma(s)(\cylinder(a_1 \dots a_n, t)) \\
%    &= \sum_{s_n \in S} \dots \sum_{s_1 \in S} \tau ^\sigma(s,a_1)(s_1) \cdots \tau^\sigma(s_{n-1},a_n)(s_n) \\
%    &\phantom{{}={}} \cdot \rho(s) \times \rho(s_1) \times \dots \times \rho(s_{n-1})(B^n_t).
%  \end{align*}
%  Hence, if we can show that
%  \[\rho(s) \times \rho(s_1) \times \dots \times \rho(s_{n-1}) (B^n_t) = (\rho(s) * \rho(s_1) * \dots * \rho(s_{n-1}))([0,t]),\]
%  the proof is done.
%  
%  The proof now proceeds by induction on the length $n$
%  of the time-bounded cylinder $\cylinder(a_1 \dots a_n, t)$.
%  If $n = 1$, then
%  \[\rho(s)(B^1_t) = \rho(s)([0,t]).\]
%  
%  If $n = k + 1$, then
%  \begin{align*}
%    &\phantom{{}={}}(\rho(s) \times \rho(s_1) \times \dots \times \rho(s_k))(B^{k+1}_t) \\
%    &= \int_0^t (\rho(s_1) \times \dots \times \rho(s_k))(B^k_{t - x}) \; \rho(s)(\wrt{x}) && \text{(Fubini)} \\
%    &= \int_0^t (\rho(s_1) * \dots * \rho(s_k))([0, t - x]) \; \rho(s)(\wrt{x}) && \text{(ind. hyp.)}\\
%    &= (\rho(s) * (\rho(s_1) * \dots * \rho(s_k)))([0, t]) && \text{(def. of convolution)}\\
%    &= (\rho(s) * \rho(s_1) * \dots * \rho(s_k))([0,t]). && \text{(associativity)} \qedhere
%  \end{align*}
%\end{proof}

Proposition \ref{prop:hyp} intuitively says that the absorption-time
of any path of length $n$ through the SMDP is distributed
as the $n$-fold convolution of its residence-time probabilities.
Therefore, the probability of doing transitions with labels $a_1, \dots, a_n$
within time $t$ is the sum of the probabilities of taking a path of length $n$
with labels $a_1, \dots, a_n$ through the SMDP,
weighted by the probability of reaching the end of each of these paths within time $t$.
This is similar in spirit to a result on phase-type distributions,
see e.g. \cite[Proposition 2.11]{pulungan2009}.

From Proposition \ref{prop:hyp} we can also derive the following
which gives a more direct inductive definition of the probability on time-bounded cylinders.
If we fix $a_1 \dots a_n$ and let $t$ vary, we get a CDF
\[\prob^\sigma(s)(a_1 \dots a_n)([0,t]) = \prob^\sigma(s)(\cylinder(a_1 \dots a_n, t)).\]

\begin{proposition}\label{prop:inductive}
  The CDF $\prob^\sigma(s)(a_1 \dots a_n)$ can be characterised inductively by
  \[\prob^\sigma(s)(a_1)([0,t]) = \sum_{s' \in S} \tau^\sigma(s,a_1)(s') \cdot \rho(s)([0,t]),\]
  \[\prob^\sigma(s)(a_1 \dots a_n)([0,t]) = \sum_{s' \in S} \tau^\sigma(s,a_1)(s') \cdot (\rho(s) * \prob^\sigma(s')(a_2 \dots a_n))([0,t]).\]
\end{proposition}
%\begin{proof}%[Proof of Proposition \ref{prop:inductive}]
%  For $n = 1$ we have
%  \[\prob^\sigma(s)(a)([0,t]) = \prob^\sigma(s)(\cylinder(a,t)) = \sum_{s' \in S} \tau^\sigma(s,a)(s') \cdot \rho(s)([0,t]).\]
%  For $n = k + 1$ we have
%  \begin{align*}
%    &\phantom{{}={}}\prob^\sigma(s)(a_1 \dots a_n) \\
%    &= \sum_{s_1 \in S} \dots \sum_{s_n \in S} \tau^\sigma(s,a_1)(s_1) \cdots \tau^\sigma(s_k, a_n)(s_n) \cdot (\rho(s) * \dots * \rho(s_n))([0,t]) \\
%    &= \sum_{s_1 \in S} \dots \sum_{s_n \in S} \tau^\sigma(s,a_1)(s_1) \cdots \tau^\sigma(s_k, a_n)(s_n) \\
%    &\phantom{{}={}}\cdot \int_0^t (\rho(s_1) * \dots * \rho(s_n))(t-x) \; \rho(s)(\wrt{x}) \\
%    &= \sum_{s_1 \in S} \tau^\sigma(s,a_1)(s_1) \cdot (\rho(s) * \prob^\sigma(s_1)(a_2 \dots a_n))([0,t]). \qedhere
%  \end{align*}
%\end{proof}

Proposition \ref{prop:inductive} also shows that our definition of faster-than
coincides with the one from \cite{PFBLM18},
except ours is reactive rather than generative.

\begin{example}\label{ex:faster-than}
  Consider the SMDPs $U$ and $V$ that are depicted in 
  Figure~\ref{fig:faster-than-c}. 
  Assuming that $F_\mu(t) \geq F_\nu(t)$ for all $t$, we now show that $U \ft V$.
  To compare $U$ and $V$, first notice that we only need to consider time-bounded cylinders of the form $\cylinder(a^n,t)$, for $n \geq 1$.
  Since the set of actions is $L = \{a\}$, the only possible valid scheduler $\sigma$ for both $U$ and $V$ is 
  the one assigning the Dirac measure $\delta_a$ to all states. We consider two cases.

  \textbf{(Case $n = 1$)} In this case we get
    \[\prob^\sigma(u_0)(\cylinder(a, t)) = F_\mu(t) \quad \text{ and } \quad \prob^\sigma(v_0)(\cylinder(a, t)) = F_\nu(t).\]
    Since we assumed $F_\mu(t) \geq F_\nu(t)$ for all $t$, this implies
    \[\prob^\sigma(u_0)(\cylinder(a, t)) \geq \prob^\sigma(v_0)(\cylinder(a, t)).\]
    
  \textbf{(Case $n > 1$)} By Proposition~\ref{prop:hyp} we have both
    \[\prob^\sigma(u_0)(\cylinder(a^n, t)) = (\mu * \nu * \eta^{*(n-2)})([0,t]) \quad \text{and} \quad \prob^\sigma(v_0)(\cylinder(a^n, t)) = (\nu * \mu * \eta^{*(n-2)})([0,t]),\] 
    where $\eta^{*n}$ is the $n$-fold convolution of $\eta$, defined inductively by 
    $\eta^{*0} = \delta_0$ and $\eta^{*(n+1)} = \eta * \eta^{*n}$. Since convolution is commutative and
    associative, and $\delta_0$ is the identity for convolution, we obtain
    \[\prob^\sigma(u_0)(\cylinder(a^n, t)) = \prob^\sigma(v_0)(\cylinder(a^n, t)).\]
    
  We therefore conclude that $U \ft V$.
\end{example}

\subsection{Comparison with Simulation and Bisimulation}
The standard notions used to compare processes are bisimulation \cite{neuhausser2007} and simulation \cite{BKHW05}.
Next we recall their definitions,
naturally extended to our setting of SMDPs.

\begin{definition}
  For an SMDP $M$, a relation $R \subseteq S \times S$ is a \emph{bisimulation relation} (resp. \emph{simulation relation}) on $M$
  if for all $(s_1, s_2) \in R$ we have
  \begin{itemize}
    \item $F_{\rho(s_1)}(t) = F_{\rho(s_2)}(t)$ (resp. $F_{\rho(s_1)}(t) \leq F_{\rho(s_2)}(t)$) for all $t \in \mathbb{R}_{\geq 0}$ and
    \item for all $a \in L$ there exists a weight function $\Delta_a : S \times S \rightarrow [0,1]$ such that
      \begin{itemize}
        \item $\Delta_a(s,s') > 0$ implies $(s,s') \in R$,
        \item $\tau(s_1,a)(s) = \sum_{s' \in S} \Delta_a(s,s')$ for all $s \in S$, and
        \item $\tau(s_2,a)(s') = \sum_{s \in S} \Delta_a(s,s')$ for all $s' \in S$.
      \end{itemize}
  \end{itemize}
  
  If there is a bisimulation relation (resp. simulation relation) $R$ such that $(s_1,s_2) \in R$,
  then we say that $s_1$ and $s_2$ are \emph{bisimilar} (resp. $s_2$ \emph{simulates} $s_1$) and write $s_1 \sim s_2$ (resp. $s_1 \simul s_2$).
\end{definition}

We lift bisimulation and simulation relations to two different SMDPs
by considering the disjoint union of the two and comparing their initial states.
We denote by $\sim$ the largest bisimulation relation and by $\simul$ the largest simulation relation.
Furthermore, we say that $U$ and $V$ are \emph{equally fast} and write $U \eqft V$
if $U \ft V$ and $V \ft U$.

\begin{example}\label{ex:comparison1}
  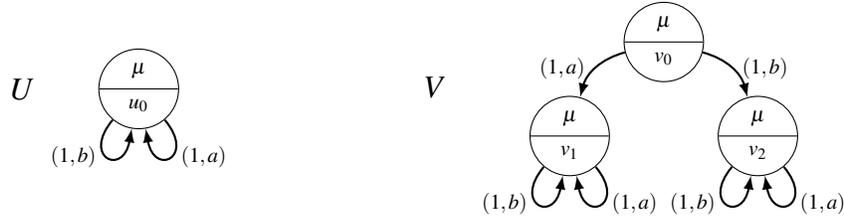
\begin{figure}
    \centering
    % First picture
    \hfill
    \begin{tikzpicture}[->, >=latex, baseline={(current bounding box.center)}]
      % Nodes
      \node[state, circle split] (0) {$\mu$ \nodepart{lower} $u_0$};
      \node[xshift=-1cm] at (0.west) {\large{$U$}};
      %
      % Edges
      \path[thick] (0) edge [out=310,in=280,looseness=6] node[right] {$(1,a)$} (0);
      \path[thick] (0) edge [out=230,in=260,looseness=6] node[left] {$(1,b)$} (0);
    \end{tikzpicture}
    \hfill
    % Second picture
    \begin{tikzpicture}[->, >=latex, baseline={(current bounding box.center)}]
      % Nodes
      \node[state, circle split] (2) [below = 2 cm of 0] {$\mu$ \nodepart{lower} $v_0$};
      \node[state, circle split] (3) [below left = 0.7cm of 2] {$\mu$ \nodepart{lower} $v_1$};
      \node[state, circle split] (4) [below right = 0.7cm of 2] {$\mu$ \nodepart{lower} $v_2$};
      \node[xshift=-2.5cm,yshift=-0.6cm] at (2.west) {\large{$V$}};
      %
      % Edges
      \path[thick] (2) edge [bend right, left] node {$(1,a)$} (3);
      \path[thick] (2) edge [bend left, right] node {$(1,b)$} (4);
      \path[thick] (3) edge [out=310,in=280,looseness=6] node[right] {$(1,a)$} (3);
      \path[thick] (3) edge [out=230,in=260,looseness=6] node[left] {$(1,b)$} (3);
      \path[thick] (4) edge [out=310,in=280,looseness=6] node[right] {$(1,a)$} (4);
      \path[thick] (4) edge [out=230,in=260,looseness=6] node[left] {$(1,b)$} (4);
    \end{tikzpicture}
    \hfill \ 
    
    \caption{Example showing that the faster-than relation and the simulation relation are incomparable.}
    \label{fig:comparison}
  \end{figure}
  
  Consider the two SMDPs $U$ and $V$ in Figure \ref{fig:comparison} with the same probability measure $\mu$ in all states.
  It is easy to see that $U$ is bisimilar to $V$, and hence $V$ also simulates $U$.
  However, we show that $U \not \ft V$ in the following way.
  Construct the scheduler $\sigma$ for $V$ by letting
  \[\sigma(v_0)(a) = 0.5, \; \sigma(v_0)(b) = 0.5, \; \sigma(v_1)(a) = 1, \, \text{ and } \, \sigma(v_2)(b) = 1.\]
  Now, for any scheduler $\sigma'$ for $U$, we must have either
  $\sigma'(u_0)(a) < 1$ or $\sigma'(u_0)(b) < 1$.
  If $\sigma'(u_0)(a) < 1$, then
  \[\sigma'(u_0)(a) > (\sigma'(u_0)(a))^2 > \dots > (\sigma'(u_0)(a))^n.\]
  Furthermore, we see that
  \[\prob^{\sigma}(v_0)(\cylinder(a^n,t)) = 0.5 \cdot \mu^{*n}(t) \text{ and } \prob^{\sigma'}(u_0)(\cylinder(a^n, t)) = (\sigma'(u_0)(a))^n \cdot \mu^{*n}(t)\]
  for $n > 1$.
  Take some $n$ such that $(\sigma'(u_0)(a))^n < 0.5$ to obtain $\prob^{\sigma'}(u_0)(\cylinder(a^n, t)) < \prob^{\sigma}(v_0)(\cylinder(a^n, t))$.
  The same procedure can be used in case $\sigma'(u_0)(b) < 1$.
  Hence we conclude that $U \not \ft V$, and therefore also that $U \not \eqft V$.
\end{example}

Example \ref{ex:comparison1} also works for schedulers with memory,
although the argument has to be modified a bit.
In that case, in each step either the probability of a trace consisting only of $a$'s
or the probability of a trace consisting only of $b$'s must decrease in $U$,
so after some number of steps, the probability of one of these two
must decrease below $0.5$, and then the rest of the argument is the same.

\begin{example}\label{ex:comparison2}
  Consider the SMDPs $U$ and $V$ in Figure \ref{fig:faster-than-c} and
  let $F_\mu = \Exp{\theta_1}$ and $F_\nu = \Exp{\theta_2}$
  be exponential distributions with rates $\theta_1 > \theta_2 > 0$.
  Then, as shown in Example \ref{ex:faster-than},
  it holds that $U \ft V$.
  However, we have both $U \not \simul V$ and $U \not \sim V$.
\end{example}

From Examples \ref{ex:comparison1} and \ref{ex:comparison2},
we get the following theorem.

\begin{theorem}\label{thm:compare}
  $\simul$ and $\ft$ are incomparable, $\sim$ and $\ft$ are incomparable, and we have ${\sim} \not \subseteq {\eqft}$.
\end{theorem}

Note that Theorem \ref{thm:compare} holds both for memoryless and for memoryful schedulers that can remember actions and states,
but it is still unclear what happens if schedulers are allowed to remember time.

%% Composition
\section{Compositionality}\label{sec:comp}
Next we introduce the notion of composition of SMDPs.
As argued in \cite{SV04}, the style of synchronous CSP composition
is the most natural one to consider for reactive probabilistic systems,
so this is the one we will adopt.
However, we leave the composition of the residence-times as a parameter,
so that we can compare different kinds of composition.

\begin{definition}
  A function $\star : \dist(\mathbb{R}_{\geq 0}) \times \dist(\mathbb{R}_{\geq 0}) \rightarrow \dist(\mathbb{R}_{\geq 0})$
  is called a \emph{residence-time composition function} if it is commutative,
  i.e. $\star(\mu,\nu) = \star(\nu,\mu)$ for all $\mu,\nu \in \dist(\mathbb{R}_{\geq 0})$.
\end{definition}

One example of such a composition function is when $\star$ is a coupling,
which is a joint probability measure such that its marginals are $\mu$ and $\nu$.
A simple special case of this is the product measure $\star(\mu,\nu) = \mu \times \nu$,
which is defined by $(\mu \times \nu)(B_1 \times B_2) = \mu(B_1) \cdot \nu(B_2)$
for all Borel $B_1$ and $B_2$.

In order to model the situation in which we want the composite system
only to take a transition when both components can take a transition,
it is natural to take the minimum of the two probabilities,
which corresponds to waiting for the slowest of the two.
In that case, we let
$F_{\star(\mu,\nu)}(t) = \min\{F_{\mu}(t), F_{\nu}(t)\}$,
and we call this \emph{minimum composition}.
Likewise, if we only require one of the components to be able to take a transition,
then it is natural to take the maximum of the two probabilities by letting
$F_{\star(\mu,\nu)}(t) = \max\{F_{\mu}(t), F_{\nu}(t)\}$,
which we call \emph{maximum composition}.
A special case of minimum composition is the composition on rates used in PEPA \cite{hillston2005},
and a special case of maximum composition is the composition on rates used in TIPP \cite{gotz1993}.

Further knowledge about the processes that are being composed
lets one define more specific composition functions.
As an example, if we know that the components only have exponential distributions,
then we can define composition functions that work directly on the rates of the distributions.
If $F_\mu = \Exp{\theta}$ and $F_{\nu} = \Exp{\theta'}$,
then one could for example let $\star(\mu,\nu)$ be such that
$F_{\star(\mu,\nu)} = \Exp{\theta \cdot \theta'}$.
This corresponds to the composition on rates that is used in SPA \cite{hermanns1998},
and we will call it \emph{product composition}.
Note that product composition is \emph{not} given by the product measure.

\begin{definition}
  Let $\star$ be a residence-time composition function.
  Then the \emph{$\star$-composition} of $U$ and $V$, denoted by $U \comp{\star} V = (S, s_0, \tau, \rho)$, is given by
  \begin{itemize}
    \item $S = U \times V$,
    \item $s_0 = (u_0, v_0)$,
    \item $\tau((u,v),a)((u',v')) = \tau_U(u,a)(u') \cdot \tau_V(v,a)(v')$ for all $a \in L$ and $(u',v') \allowbreak \in S$, and
    \item $\rho((u,v)) = \star(\rho_U(u), \rho_V(v))$. \qedhere
  \end{itemize}
%  for all $(u,v) \in S$.
\end{definition}

We write $u \comp{\star} v$
to denote the composite state $(u,v)$ of $U \comp{\star} V$
where $u \in S_U$ and $v \in S_V$.

%%%%%%%%%%%%%%%%%%%%%%%%%%%%%%%%%%%%%%%%%%%%%%%
%% TIMING ANOMALIES
%%%%%%%%%%%%%%%%%%%%%%%%%%%%%%%%%%%%%%%%%%%%%%%
\subsection{Parallel Timing Anomalies}\label{sec:anomalies}
If we  have two components $U$ and $V$,
and we know that $U$ is faster than $V$,
then if $V$ is in parallel with some context $W$,
we would expect this composition to become faster when
we replace the component $V$ with the component $U$.
However, sometimes this fails to happen,
and we will call such an occurrence a \emph{parallel timing anomaly}.

In this section we show that parallel timing anomalies can occur
for some of the kinds of composition discussed in Section \ref{sec:comp}.
We do this by giving different contexts $W$ for the SMDPs $U$ and $V$
from Figure \ref{fig:faster-than-c},
for which it was shown in Example \ref{ex:faster-than} that $U \ft V$.
Our examples of parallel timing anomalies make no use of non-determinism or probabilistic branching,
thus showing that the parallel timing anomalies are caused
inherently by the timing behaviour of the SMDPs.
For ease of presentation, we let the set of labels $L$
consist only of the label $a$ in this section.

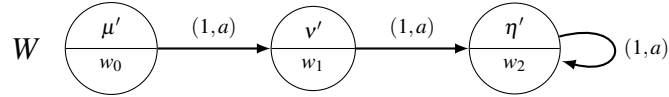
\begin{figure}
  \centering
  \hfill
  \begin{tikzpicture}
    %Nodes
    \node[state, circle split] (0) {$\mu'$ \nodepart{lower} $w_0$};
    \node[state, circle split] (1) [right = 1.5cm of 0]{$\nu'$ \nodepart{lower} $w_1$};
    \node[state, circle split] (8) [right = 1.5cm of 1]{$\eta'$ \nodepart{lower} $w_2$};
    
    \node[xshift=-0.5cm] at (0.west) {\large{$W$}};
    
    % Edges
    \path[thick] (0) edge [above] node {$(1,a)$} (1);
    \path[thick] (1) edge [above] node {$(1,a)$} (8);
    \path[thick, loop right] (8) edge [right] node {$(1,a)$} (8);
  \end{tikzpicture}
  \hfill \
  \caption{For different instantiations of $\mu'$, $\nu'$, and $\eta'$, the context $W$ leads to parallel timing anomalies for product, minimum, and maximum rate composition, respectively.}
  \label{fig:anomalies}
\end{figure}

Consider the two SMDPs $U$ and $V$ depicted in Figure \ref{fig:faster-than-c}
and let the context $W$ be given by Figure \ref{fig:anomalies}.
For the examples in this section,
let $F_\mu = \Exp{2}$, $F_\nu = \Exp{0.5}$,
and let $\eta = \eta'$ be arbitrary.
From Example \ref{ex:faster-than} we know that $U \ft V$.

\begin{example}[Product composition]\label{ex:prodanomaly}
  Let $\star$ be product composition
  and let $F_{\mu'} = \Exp{10}$ and $F_{\nu'} = \Exp{0.1}$.
  In $U \comp{\star} W$, the rates in the first two states will then be $20$ and $0.05$,
  and in $V \comp{\star} W$ they will be $5$ and $0.2$.
  Consider the time-bounded cylinder $\cylinder(aa, 2)$.
  Then we see that
  \[\prob(u_0 \comp{\star} w_0)(\cylinder(aa, 2)) \approx 0.09 \quad \text{and} \quad \prob(v_0 \comp{\star} w_0)(\cylinder(aa, 2)) \approx 0.30,\]
  showing that $U \comp{\star} W \not\ft V \comp{\star} W$.
  Hence we have a parallel timing anomaly.
  What happens is that in the process $V \comp{\star} W$ the probability of taking a transition before time $2$
  with rate $5$ is already very close to $1$,
  so the process $U \comp{\star} W$ does not gain much by having a rate of $20$,
  whereas in the next step, $V \comp{\star} W$ gains a lot of probability by having a rate of $0.5$
  compared to the rate $0.05$ of $U \comp{\star} W$.
\end{example}

\begin{example}[Minimum composition]\label{ex:minanomaly}
  Let $\star$ be minimum composition
  and let $F_{\mu'} = \Exp{1}$ and $F_{\nu'} = \Exp{2}$.
  The rates of $U \comp{\star} W$ are then $1$ and $0.5$,
  whereas they are $0.5$ and $2$ in $V \comp{\star} W$.
  Then we get
  \[\prob(u_0 \comp{\star} w_0)(\cylinder(aa, 2)) \approx 0.40 \quad \text{and} \quad \prob(v_0 \comp{\star} w_0)(\cylinder(aa, 2)) \approx 0.51,\]
  so $U \comp{\star} W \not\ft V \comp{\star} W$.
  What happens in this example is that in the second step,
  $U \comp{\star} W$ has the same rate as $V \comp{\star} W$ had in the first step.
  This means that $U \comp{\star} W$ must be proportionally faster in the second step.
  However, $V \comp{\star} W$ has a rate of $2$ in the second step,
  but $U \comp{\star} W$ only had a rate of $1$ in the first step.
\end{example}

\begin{example}[Maximum composition]\label{ex:maxanomaly}
  Let $\star$ be maximum composition
  and let $F_{\mu'} = \Exp{2}$ and $F_{\nu'} = \Exp{1}$.
  $U \comp{\star} W$ then has rates $2$ and $1$,
  and $V \comp{\star} W$ has rates $2$ and $2$.
  Then
  \[\prob(u_0 \comp{\star} w_0)(\cylinder(aa, 2)) \approx 0.75 \quad \text{and} \quad \prob(v_0 \comp{\star} w_0)(\cylinder(aa, 2)) \approx 0.91,\]
  so $U \comp{\star} W_3 \not\ft V \comp{\star} W_3$.
  The reason for the timing anomaly in this case is clear:
  $V \comp{\star} W$ simply has a higher rate in each step than $U \comp{\star} W$ does.
\end{example}

%%%%%%%%%%%%%%%%%%%%%%%%%%%%%%%%%%%%%%%%%%%%%%%
%% AVOIDING TIMING ANOMALIES
%%%%%%%%%%%%%%%%%%%%%%%%%%%%%%%%%%%%%%%%%%%%%%%
\subsection{Avoiding Parallel Timing Anomalies}
We have seen in the previous section that parallel timing anomalies can occur.
We now wish to understand what kind of contexts do not lead to timing anomalies.
In this section we assume that the set $L$ of transition labels is a finite set.
Also, we fix a residence-time composition function $\star$
and two additional SMDPs $W = (S_W, w_0, \tau_W, \rho_W)$ and $W' = (S_{W'}, w'_{0}, \tau_{W'}, \rho_{W'})$
which should be thought of as contexts.
Next we identify conditions on $(W,w_0)$ such that
$U \ft V$ will imply $U \comp{\star} W \ft V \comp{\star} W$.

We first give conditions that over-approximate the faster-than relation between the composite systems
by requiring that when $U$ and $W$ are put in parallel,
then the composite system is point-wise faster than $U$ along all paths.
Likewise, we require that when $V$ and $W$ are put in parallel,
the composite system is point-wise slower than $V$ along all paths.
If we already know that $U$ is faster than $V$,
this will imply by transitivity that $U \comp{\star} W$ is faster than $V \comp{\star} W$.
We have already seen in Example \ref{ex:faster-than}
that a process $U$ need not be point-wise faster than $V$ along all paths
in order for $U$ to be faster than $V$.
However, by imposing this condition,
we do not need to compare convolutions of distributions,
but can compare the distributions directly.

We will say that an SMDP $M$ has a \emph{deterministic Markov kernel}
if for all states $s$ and labels $a$,
there is at most one state $s'$ such that $\tau(s,a)(s') > 0$.
A \emph{state path} of length $n$ in $M$ is a sequence $s_1 \dots s_n$
such that $s_1$ is the initial state and for all $i$ there exists $a \in L$ such that $\tau(s_i, a)(s_{i+1}) > 0$.

%\begin{definition}
%  A \emph{state path in $M$} is a sequence of states $s_1,s_2, \dots$
%  where for all $i \in \mathbb{N}$ there exists a label $a \in L$
%  such that $\tau(s_i, a)(s_{i+1}) > 0$.
%  For a state path $\pi = s_1, s_2, \dots$, we let $\pi\sproj{i} = s_i$,
%  $\pi|^i = s_i, s_{i+1}, \dots$, $\pi|_i = s_1, s_2, \dots, s_i$,
%  and we let $\spaths{}{M}$ denote the set of all state paths in $M$.
%  For a state $s \in S$, we let $\spaths{}{s} = \{ \pi \in \spaths{}{M} \mid \pi\sproj{1} = s\}$
%  and we let $\spaths{n}{s} = \{\pi|_n \mid \pi \in \spaths{}{s}\}$.
%\end{definition}

%By using the notion of a state path, we can write Proposition \ref{prop:hyp} as
%\begin{align*}
%  &\phantom{{}={}}\prob^\sigma(s)(\cylinder(a_1 \dots a_n, x)) \\
%  &= \sum_{\pi \in \spaths{n+1}{s}} \tau^\sigma(\pi\sproj{1},a_1)(\pi\sproj{2}) \cdots \tau^\sigma(\pi\sproj{n},a_n)(\pi\sproj{n+1}) \cdot (\rho(\pi\sproj{1}) * \dots * \rho(\pi\sproj{n}))([0,x]).
%\end{align*}

\begin{definition}\label{def:safe}
  Let $n \in \mathbb{N}$.
  We say that $\star$ is \emph{$n$-monotonic in $U$, $V$, $W$, and $W'$},
  written $(U,W) \mon{n}{\star} (V,W')$,
  if $W'$ has a deterministic Markov kernel and for all state paths $u_1 \dots u_n$, $v_1 \dots v_n$, $w_1 \dots w_n$, and $w'_1 \dots w'_n$ we have
  \begin{itemize}
    \item $F_{\rho(u_i \comp{\star} w_i)}(t) \geq F_{\rho_U(u_i)}(t)$ and
      $F_{\rho_V(v_i)}(t) \geq F_{\rho(v_i \comp{\star} w'_i)}(t)$ for all $t \in \mathbb{R}_{\geq 0}$ and $1 \leq i \leq n$,
    \item for all schedulers $\sigma_U$ for $U$ there exists a scheduler $\sigma_{U,W}$ for $U \comp{\star} W$ such that we have
      \[\tau^{\sigma_{U,W}}(u_i \comp{\star} w_i, a)(u_{i+1} \comp{\star} w_{i+1}) \geq \tau^{\sigma_U}_U(u_i,a)(u_{i+1}), \quad \text{and}\]
    \item for all schedulers $\sigma_{V,W'}$ for $V \comp{\star} W'$ there exists a scheduler $\sigma_V$ for $V$ such that we have
      \[\tau^{\sigma_V}_V(v_i,a)(v_{i+1}) \geq \tau^{\sigma_{V,W'}}(v_i \comp{\star} w'_i,a)(v_{i+1} \comp{\star} w'_{i+1})\]
  \end{itemize}
  and for all $a \in L$ and $1 \leq i < n$.
  Furthermore, we will say that $\star$ is \emph{monotonic in $U$, $V$, $W$, and $W'$}
  and write $(U,W) \mon{}{\star} (V,W')$,
  if it is $n$-monotonic in $U$, $V$, $W$, and $W'$ for all $n \in \mathbb{N}$.
\end{definition}

Clearly, if $(U,W) \mon{n}{\star} (V,W')$,
then $(U,W) \mon{m}{\star} (V,W')$ for all $m \leq n$.
The next result shows that if $(U,W) \mon{}{\star} (V,W')$,
then we are guaranteed to avoid parallel timing anomalies.

\begin{theorem}\label{thm:avoid}
  If $(U,W) \mon{}{\star} (V,W')$ as well as $U \ft V$ and $W \ft W'$,
  then we have $U \comp{\star} W \ft V \comp{\star} W'$.
\end{theorem}

The special case where $W = W'$ shows that this
condition is sufficient to avoid parallel timing anomalies.
We do not know if it is decidable whether $(U,W) \mon{}{\star} (V,W')$.
However, there is a stronger condition which is decidable
in the case of finite SMDPs.
We present it in the next definition.

\begin{definition}\label{def:smon}
  We say that $\star$ is \emph{strongly} $n$-monotonic in $U$, $V$, $W$, and $W'$
  and write $(U,W) \smon{n}{\star} (V,W')$
  if $W'$ has a deterministic Markov kernel and for all state paths $u_1 \dots u_n$, $v_1 \dots v_n$, $w_1 \dots w_n$, and $w'_1 \dots w'_n$,
  \begin{itemize}
    \item $F_{\rho(u_i \comp{\star} w_i)}(t) \geq F_{\rho_U(u_i)}(t)$ and
      $F_{\rho_V(v_i)}(t) \geq F_{\rho(v_i \comp{\star} w'_i)}(t)$ for all $t \in \mathbb{R}_{\geq 0}$ and $1 \leq i \leq n$
    \item for all schedulers $\sigma_U$ for $U$ and all schedulers $\sigma_{U,W}$ for $U \comp{\star} W$ we have
      \[\tau^{\sigma_{U,W}}(u_i \comp{\star} w_i,a)(u_{i+1} \comp{\star} w_{i+1}) \geq \tau^{\sigma_U}_U(u_i,a)(u_{i+1}), \quad \text{and}\]
    \item for all schedulers $\sigma_{V,W'}$ for $V \comp{\star} W'$ and all schedulers $\sigma_V$ for $V$ we have
      \[\tau^{\sigma_V}_V(v_i,a)(v_{i+1}) \geq \tau^{\sigma_{V,W'}}(v_i \comp{\star} w'_i,a)(v_{i+1} \comp{\star} w'_{i+1})\]
  \end{itemize}
  for all $a \in L$ and $1 \leq i < n$.
  If $(U,W) \smon{n}{\star} (V,W')$ for all $n \in \mathbb{N}$,
  we say that $\star$ is \emph{strongly monotonic} in $U$, $V$, $W$, and $W'$
  and write $(U,W) \smon{}{\star} (V,W')$.
\end{definition}

The conditions of Definition \ref{def:smon} are
similar to the conditions from Definition \ref{def:safe},
but the existential quantifier is strengthened to a universal quantifier.
It is obvious that $(U,W) \smon{}{\star} (V,W')$ implies $(U,W) \mon{}{\star} (V,W')$,
and hence we get the following corollary.

\begin{corollary}
  If $(U,W) \smon{}{\star} (V,W')$ as well as $U \ft V$ and $W \ft W'$,
  then $U \comp{\star} W \ft V \comp{\star} W'$.
\end{corollary}

\begin{example}
  Let $U$ and $V$ be given by Figure \ref{fig:faster-than-c}
  with $F_\mu \geq F_\nu$ as in Example \ref{ex:faster-than}.
  Let $\star$ be minimum rate composition and consider the context $W$ from Figure \ref{fig:anomalies},
  where $\mu' = \mu$, $\nu' = \nu$, and $\eta' = \eta$.
  There is only one possible scheduler $\sigma$, which is the Dirac measure at $a$,
  and hence it is clear that the second and third conditions are satisfied. We also find that
  \begin{align*}
    F_{\rho(u_0 \comp{\star} w_0)}(t) = F_{\rho_U(u_0)}(t) & \quad & F_{\rho_V(v_0)}(t) = F_{\rho(v_0 \comp{\star} w_0)}(t) \\
    F_{\rho(u_1 \comp{\star} w_1)}(t) = F_{\rho_U(u_1)}(t) & \quad & F_{\rho_V(v_1)}(t) = F_{\rho(v_1 \comp{\star} w_1)}(t) \\
    F_{\rho(u_2 \comp{\star} w_2)}(t) = F_{\rho_U(u_2)}(t) & \quad & F_{\rho_V(v_2)}(t) = F_{\rho(v_2 \comp{\star} w_2)}(t)
  \end{align*}
  and hence the first condition is also satisfied, so $(U,W) \smon{}{\star} (V,W)$.
\end{example}

\begin{example}
  All the examples we gave in Section \ref{sec:anomalies} fail to be monotonic,
  and hence also fail to be strongly monotonic, since they all violate the first condition of monotonicity.
  
  In Example \ref{ex:prodanomaly}, this is because
  \[F_{\rho(v_0)}(t) = \Exp{0.5}(t) < \Exp{5}(t) = F_{\rho(v_0 \comp{\star} w_0)}(t) \quad \text{for any $t > 0$.}\]
  Likewise, in Example \ref{ex:minanomaly} we have
  \[F_{\rho(u_0 \comp{\star} w_0)}(t) = \Exp{1}(t) < \Exp{2}(t) = F_{\rho(u_0)}(t) \quad \text{for any $t > 0$.}\]
  Finally, in Example \ref{ex:maxanomaly} we have
  \[F_{\rho(v_0)}(t) = \Exp{0.5}(t) < \Exp{2}(t) = F_{\rho(v_0 \comp{\star} w_0)}(t) \quad \text{for any $t > 0$.} \qedhere\]
\end{example}

The following is the main theorem of our paper,
showing that strong monotonicity implies the absence of parallel timing anomalies.

\begin{theorem}
  If $(U,W) \smon{}{\star} (V,W')$ as well as $U \ft V$ and $W \ft W'$,
  then $U \comp{\star} W \ft V \comp{\star} W'$.
\end{theorem}

We now wish to show that it is decidable whether $(U,W) \smon{}{\star} (V,W')$ for finite SMDPs,
thereby giving a decidable condition for avoiding timing anomalies.
To do this, we first show that in order to establish strong monotonicity,
it is enough to consider paths up to length
\[m = \max\{|S_U| \cdot |S_W|, |S_V| \cdot |S_{W'}|\} + \max\{|S_U|,|S_V|,|S_W|,|S_{W'}|\} + 1,\]
due to the fact that they start repeating,
as can be shown by a simple pigeonhole argument.

%\begin{lemma}\label{lem:cycles}
%  Let $U$ and $V$ be two finite, pointed SMDPs.
%  For any state paths $\pi_U$ and $\pi_V$
%  of length $l > |S_U| \cdot |S_V|$,
%  there will be $i < j \leq |S_U| \cdot |S_V| + 1$ such that
%  $\pi_U\sproj{i} = \pi_U\sproj{j}$, $\pi_V\sproj{i} = \pi_V\sproj{j}$.
%\end{lemma}
%\begin{proof}
%  Since there are $|S_U| \cdot |S_V|$ ways of choosing a pair $(u_i,v_j) \in S_U \times S_V$
%  of states from $U$ and $V$, if we pair the states of $\pi_U$ and $\pi_V$
%  such that we get the pairs $(\pi_U\sproj{1},\pi_V\sproj{1})$, $(\pi_U\sproj{2},\pi_V\sproj{2})$, $\dots$, $(\pi_U\sproj{l},\pi_V\sproj{l})$,
%  there must be two of these pairs that are the same because $l > |S_U| \cdot |S_V|$.
%  Hence we get states $\pi_U\sproj{i} = \pi_U\sproj{j}$
%  and $\pi_V\sproj{i} = \pi_V\sproj{j}$ with $i < j \leq n$.
%  It also follows that $i$ and $j$ can be chosen so that $i < j \leq |S_U| \cdot |S_V|$,
%  because otherwise we would have $j - i > |S_U| \cdot |S_V|$ different pairs
%  \[(\pi_U\sproj{i},\pi_V\sproj{i}), (\pi_U\sproj{i+1},\pi_V\sproj{i+1}), \dots, (\pi_U\sproj{j},\pi_V\sproj{j}),\]
%  contradicting the fact that there are only $|S_U| \cdot |S_V|$ such different pairs.
%\end{proof}

\begin{lemma}\label{lem:m-safe}
  Let $U$, $V$, $W$, and $W'$ be finite.
  If $(U,W) \smon{m}{\star} (V,W')$, then $(U,W) \smon{}{\star} (V,W')$.
\end{lemma}

We can now use the first-order theory of the reals to show
that strong monotonicity is a decidable property.
In order for this to work, we must be able to check the first condition of strong monotonicity
in the first-order theory of the reals.
We therefore need the sets of $t \in \mathbb{R}_{\geq 0}$ in that condition
to be expressible in the first-order theory of reals,
which can be accomplished by assuming that these sets are semialgebraic.

\begin{theorem}\label{thm:decidable}
  Let $U$, $V$, $W$, and $W'$ be finite.
  If for all state paths $u_1 \dots u_n$, $v_1 \dots v_n$, $w_1 \dots w_n$, and $w'_1 \dots w'_n$
  we have that $\{t \in \mathbb{R}_{\geq 0} \mid F_{\rho(u_i \comp{\star} w_i)}(t) \geq F_{\rho_U(u_i)}(t)\}$
  and $\{t \in \mathbb{R}_{\geq 0} \mid F_{\rho_V(v_i)}(t) \geq F_{\rho(v_i \comp{\star} w'_i)}(t)\}$
  are semialgebraic sets for all $1 \leq i \leq m$, then it is decidable whether $(U,W) \smon{}{\star} (V,W')$.
\end{theorem}
%\begin{proof}%[Proof of Theorem \ref{thm:decidable}]
%  Note first of all that since $L$ and $W'$ are finite,
%  it is decidable whether $W'$ has a deterministic Markov kernel by looking at all the states.
%  By Lemma \ref{lem:m-safe}, it suffices to check
%  whether $(U,W) \smon{m}{\star} (V,W')$ where
%  \[m = \max\{|S_U|\cdot|S_W|,|S_V|\cdot|S_{W'}|\} + \max\{|S_U|,|S_V|,|S_W|,|S_{W'}|\} + 1.\]
%  This can be done by exploiting the decidability
%  of the first-order theory of the reals in the following way.
%  Since $L$ is finite and $U$, $V$, $W$, and $W'$ are all finite,
%  there are finitely many state paths $\pi_U \in \spaths{m}{u_0}$,
%  $\pi_V \in \spaths{m}{v_0}$, $\pi_W \in \spaths{m}{w_0}$, and $\pi_{W'} \in \spaths{m}{w'_0}$.
%  Because of this, and since the sets
%  \[\{t \in \mathbb{R}_{\geq 0} \mid F_{\rho(\pi_U\sproj{i} \comp{\star} \pi_W\sproj{i})}(t) \geq F_{\rho_U(\pi_U\sproj{i})}(t)\}\]
%  and
%  \[\{t \in \mathbb{R}_{\geq 0} \mid F_{\rho_V(\pi_V\sproj{i})}(t) \geq F_{\rho(\pi_V\sproj{i} \comp{\star} \pi_{W'}\sproj{i})}(t)\},\]
%  which we need to check for the first condition,  
%  were assumed to be semialgebraic,
%  it is possible to express the conditions of Definition \ref{def:smon}
%  in the first-order theory of the reals,
%  using finitely many quantifiers and inequalities.
%  Since the first-order theory of the reals is decidable,
%  the truth value of the resulting formula is decidable.
%\end{proof}

For uniform and exponential distributions with minimum or maximum composition,
the corresponding sets are all semialgebraic,
and the same is true for exponential distributions with product composition.
Theorem \ref{thm:decidable} can therefore be used for these types of composition.

It turns out that strong monotonicity implies the absence of non-determinism.

\begin{proposition}\label{prop:singleton}
  If $(U,W) \smon{}{\star} (V,W')$, then $L$ is a singleton set or $u_0$ is a deadlock state,
  i.e. the transition probability is zero from $u_0$ to any other state.
\end{proposition}

However, strong monotonicity still makes sense as a condition,
since all our examples of timing anomalies in Section \ref{sec:anomalies}
have no non-determinism.

%\begin{proof}[Proof of Proposition \ref{prop:singleton}]
%  We prove the contrapositive.
%  Suppose $|L| > 1$ and $u_0$ is not a deadlock state.
%  Because $u_0$ is not a deadlock state,
%  there must exist some state path $\pi_U$ such that
%  $\pi_U\sproj{1} = u_0$ and $\tau_U(\pi_U\sproj{1},a)(\pi_U\sproj{2}) > 0$ for some $a \in L$.
%  Since $|L| > 1$, we can find some $b \in L$ with $a \neq b$.
%  Now construct schedulers given by $\sigma_{U,W}(s) = \delta_b$ and $\sigma_U(s) = \delta_a$
%  for any state $s$.
%  Then
%  \[\tau^{\sigma_{U,W}}(\pi_U\sproj{1} \comp{\star} \pi_W\sproj{1}, a)(\tau_U\sproj{2} \comp{\star} \tau_W\sproj{2}) = 0\]
%  but
%  \[\tau_U^{\sigma_U}(\pi_U\sproj{1}, a)(\tau_U\sproj{2}) > 0,\]
%  and hence the first condition of Definition \ref{def:smon}
%  is violated.
%\end{proof}

%% Conclusion
\section{Conclusion}
In this paper, we have investigated the notion of a process being faster than another process in the context of semi-Markov decision processes.
We have given a trace-based definition of a faster-than relation,
and shown how this notion relates to the usual notions of simulation and bisimulation
and to convolutions of distributions.
By considering composition as being parametric in how the residence times of states are combined,
we have given examples showing that our faster-than relation gives rise to parallel timing anomalies
for many of the popular ways of composing rates.
We have therefore given sufficient conditions for how such parallel timing anomalies can be avoided,
and we have shown that these conditions are decidable.

While the conditions for strong monotonicity are so strict that they rule out non-determinism,
it seems difficult to obtain better results that are still decidable.
This is because the faster-than relation itself is undecidable whenever non-determinism is introduced,
and even without non-determinism, it has close connections to an important problem in number theory \cite{PFBLM18} whose decidability status is a long-standing open problem.
Because of this, an interesting future work direction is to explore the boundaries of decidability 
for conditions to avoid timing anomalies.

Furthermore, the conditions we have given for avoiding timing anomalies
do not look at the context in isolation,
but depend also on the processes that are being swapped.
It would be preferable to have conditions on a context
that would guarantee the absence of parallel timing anomalies
no matter what processes are being swapped.

\paragraph{Acknowledgements.}
Mathias R.\ Pedersen was supported by the ASAP project (no.\ 4181-00360) funded by the Danish Council for Independent Research. Kim G.\ Larsen was supported by the ERC Advanced Grant LASSO (no.\ 867096).

%\nocite{*}
\bibliographystyle{eptcs}
\bibliography{bibliography}

\end{document}